\documentclass[12pt]{iopart}
\begin{document}
\title[Quasi-isotropic solution near a cosmological singularity for a two-fluid model]
{Quasi-isotropic solution of the Einstein equations near
a cosmological singularity for a two-fluid cosmological model}
\author{I.M. Khalatnikov$^{1,2,3}$, A.Yu. Kamenshchik$^{1,2}$,  
M. Martellini$^{2,4,5}$ \\and A.A. Starobinsky$^{1}$}
\address{1\ L.D. Landau Institute for Theoretical Physics, 
Russian Academy of Sciences, Kosygin str. 2, 119334, Moscow, Russia} 
\address{2\ Landau Network - Centro Volta, Villa Olmo, via Cantoni 1, 22100 
Como, Italy}
\address{3\ Tel Aviv University,
Tel Aviv University, Raymond and Sackler
Faculty of Exact Sciences, School of Physics and Astronomy,
Ramat Aviv, 69978, Israel}
\address{4\ Dipartimento di Scienze Matematiche, Fisiche e Chimiche, Universit\`a
dell'Insubria, via Valleggio 11, 22100 Como, Italy}
\address{5\ Sez. INFN di Milano, Italy}
\date{}
\maketitle
\begin{abstract}
The quasi-isotropic inhomogeneous solution of the Einstein equations 
near a cosmological singularity in the form of a series expansion in
the synchronous system of reference, first found by Lifshitz and 
Khalatnikov in 1960, is generalized to the case of a two-fluid 
cosmological model. This solution describes non-decreasing modes of 
adiabatic and isocurvature scalar perturbations and gravitational 
waves in the regime when deviations of a space-time metric from the 
homogeneous isotropic Friedmann-Robertson-Walker (FRW) background are 
large while locally measurable quantities like Riemann tensor 
components are still close to their FRW values. The general structure 
of the perturbation series is presented and the first coefficients of 
the series expansion for the metric tensor and the fluid energy 
densities and velocities are calculated explicitly.
\end{abstract}

\section{Introduction}

In the recent paper \cite{we}, we presented the generalization of the 
quasi-isotropic solution of the Einstein equations near a cosmological 
singularity found by Lifshitz and Khalatnikov \cite{Lif-Khal} for the 
Universe filled by radiation  with the equation of state $p = 
\frac{\varepsilon}{3}$ to the case of an arbitrary one-fluid 
cosmological model. Here, this solution is further generalized to the 
case of the Universe filled by two ideal barotropic fluids.
 
As is well known, modern cosmology deals with many very different 
types of matter. In comparison with the old standard model of the hot 
Universe (the Big Bang), the situation has been dramatically changed, 
first, with the development of inflationary cosmological models which 
contain an inflaton effective scalar field or/and other exotic types 
of matter as an important ingredient \cite{inflation}, and second, with 
the understanding that the main part of the non-relativistic matter in 
the present Universe is non-baryonic -- cold dark matter (CDM). 
Furthermore, the appearance of brane and M-theory cosmological models~
\cite{brane} and the discovery of the cosmic acceleration~ 
\cite{cosmic} (see also \cite{varun} for a review) suggests that 
matter playing an essential role at different stages of cosmological 
evolution is multi-component generically, and these components may 
obey very different equations of state.  Moreover, the very notion of 
the equation of state appears to be not fundamental; it has only a 
limited range of validity as compared to a more fundamental 
field-theoretical description. From this general point of view, the 
generalization of the quasi-isotropic solution to the case of two 
ideal barotropic fluids with constant but different $p/{\varepsilon}$ 
ratios seems to be a natural and important next logical step. The most 
straightforward candidates for these fluids are non-relativistic  
matter (CDM) with $p=0$ and radiation. One more popular candidate is 
ultra-stiff matter ($p=\varepsilon$) which underlying physical model 
is a minimally coupled to gravity scalar field in the regime when its 
potential energy may be neglected as compared to its kinetic energy~
\cite{BKh72} (in particular, it may be the inflaton field itself after 
the end of inflation).

To explain the physical sense of the quasi-isotropic solution, let us 
remind that it represents the most generic spatially inhomogeneous 
generalization of the FRW space-time in which the space-time is locally
FRW-like near the cosmological singularity $t=0$ (in particular, its
Weyl tensor is much less than its Riemann tensor). On the other hand, 
generically it is very inhomogeneous globally and may have a very 
complicated spatial topology. As was shown in \cite{Lif-Khal1,we} (see 
also \cite{Deruelle}), such a solution contains 3 arbitrary functions 
of space coordinates. From the FRW point of view, these 3 degrees of 
freedom represent the growing (non-decreasing in terms of metric 
perturbations) mode of adiabatic perturbations and the non-decreasing 
mode of gravitational waves (with two polarizations) in 
the case when deviations of a space-time metric from the FRW one are 
not small. So, the quasi-isotropic solution is not a generic solution 
of the Einstein equations with a barotropic fluid. Therefore, one 
should not expect this solution to arise in the course of generic 
gravitational collapse (in particular, inside a black hole event 
horizon). The generic solution near a space-like curvature singularity 
(for $p<\varepsilon$) has a completely different structure consisting 
of the infinite sequence of anisotropic vacuum Kasner-like eras with 
space-dependent Kasner exponents \cite {BLH}.

 For this reason, the quasi-isotropic solution had not attracted 
much interest for about twenty years. Its new life began after the 
development of successful inflationary models (i.e., with "graceful 
exit" from inflation) and the theory of generation of perturbations
during inflation, because it had immediately become clear that 
generically (without fine tuning of initial conditions) scalar metric
perturbations after the end of inflation remained small in a finite 
region of space which was much less than the whole causally connected 
space volume produced by inflation. It appears that the 
quasi-isotropic solution can be used for a global description of a 
part of space-time after inflation which belongs to "one 
post-inflationary universe". The latter is defined as a connected part 
of space-time where the hyper-surface $t=t_f({\bf r})$ describing the 
moment when inflation ends is space-like and, therefore, can be made 
the surface of constant (zero) synchronous time by a coordinate 
transformation. This directly follows from the derivation of 
perturbations generated during inflation given in \cite{Star82} (see 
Eq. (17) of that paper) which is valid in case of large perturbations, 
too. Thus, when used in this context, the quasi-isotropic solution 
represents an {\em intermediate} asymptotic regime during expansion of 
the Universe after inflation. The synchronous time $t$ appearing in it
is the proper time since the {\em end} of inflation, and the region 
of validity of the solution is from $t=0$ up to a moment in future 
when spatial gradients become important. For sufficiently large 
scales, the latter moment may be rather late, even of the order or
larger than the present age of the Universe. Note also  the analogue 
of the quasi-isotropic solution {\em before} the end of inflation is 
given by the generic quasi-de Sitter solution found in \cite{Star83}.
Both solutions can be smoothly matched across the hypersurface of 
the end of inflation.

Generalization of the quasi-isotropic solution to the case of multiple 
fluids provides a possibility to investigate the growing isocurvature 
mode of scalar perturbations and its effect on the growing mode of 
adiabatic perturbations in the regime when a space-time metric 
{\em may not} be globally represented as the FRW metric with small 
perturbations. Isocurvature perturbation of the energy density of
a subdominant fluid need not be small, too. Note that the splitting
of scalar perturbations into the adiabatic and the isocurvature mode
is, to some extent, conventional (especially, in models with decay
of one forms of matter into others, e.g., in the so called "curvaton" 
model \cite{LW02}). Of course, this does not affect theoretical
predictions for any observable quantity since contributions from all
modes should be taken into account. We use the most natural splitting 
which is essentially the same as in the standard cosmological 
($\Lambda$)CDM + radiation model, namely, that metric perturbations
for the growing isocurvature mode are zero at $t=0$. However, they are 
not zero for $t>0$ and grow with time. 

It is well known that the growing adiabatic mode remains constant in 
the synchronous reference system in the leading long-wave
approximation for any one-fluid (and even any one-component matter) 
model both in the linear regime and with all back reaction effects 
taken into account. This immediately follows from the very form of 
the quasi-isotropic solution for one fluid -- it corresponds to
the {\em factorization} of spatial and temporal variables in its
leading term. In a multi-component case, this property is known to 
be valid in the linear approximation, namely, that there always 
exists one solution of equations for perturbations (which we just 
call the growing adiabatic mode) which remains constant in the 
long-wave regime (see, e.g., \cite{STY01} and for a more recent 
discussion \cite{MWU02}). However, some concern was expressed (e.g., 
in \cite{GB02}) if this remains valid in the long-wave 
non-linear multi-component regime. We will show that the 
factorization of variables in the first term occurs in the
generalized quasi-isotropic solution, too, and this term remains 
the leading one either until spatial gradients become important, 
or until the energy density of a firstly subdominant fluid 
becomes comparable to that of the dominant one (that results in 
the change of the expansion law).

\section{Series expansion of the quasi-isotropic solution for a 
two-fluid case}

Let us consider a FRW model with two barotropic fluids 
satisfying the equations of state:
\begin{equation}
p_l = k_l \varepsilon_l,~~l=1,2
\label{state1}
\end{equation}
where $p_l$ and $\varepsilon_l$ are the pressure and the energy 
density of the corresponding fluid. For usual fluids, 
$0\le k_l\le 1$. However, similar to the case of one fluid, our
solution remains valid for a larger range $k_l> - {1\over 3}$ for
both fluids. Let us take
\begin{equation}
k_2 > k_1.
\label{inequality}
\end{equation}

As usually \cite{Lif-Khal1,Land-Lif}, we will work in the 
synchronous system of reference 
\begin{equation}
ds^2 = dt^2 - \gamma_{\alpha\beta}dx^{\alpha}dx^{\beta}.
\label{sinchron}
\end{equation}
Though this system of reference is not fixed uniquely (in other words,
some of arbitrary functions entering into a solution written in this
system are fictitious, or gauge artifacts), the choice of the initial
cosmological singularity at the hypersurface $t=0$ in the 
quasi-isotropic solution (see Eq. (\ref{quasi}) below) removes half 
of this ambiguity leaving only 3D spatial rotations in the leading 
term (with corresponding terms in higher orders) as the remaining 
gauge freedom. 

To find the time dependence of the leading term of the quasi-isotropic 
solution,  we write down the FRW energy equation in the vicinity of 
the cosmological singularity which reads
\begin{equation}
\frac{\dot{a}^2}{a^2} + \frac{k}{a^2} = \frac{A_1}{a^{\alpha_1}} +  
\frac{A_2}{a^{\alpha_2}}~
\label{Friedmann}
\end{equation}
where $A_1$ and $A_2$ are some constants, while the exponents 
$\alpha_1$ and $\alpha_2$ are defined as 
\begin{equation}
\alpha_1 = 3(1+k_1),
\label{alpha1}
\end{equation}
\begin{equation}
\alpha_2 = 3(1+k_2) > \alpha_1.
\label{alpha2}
\end{equation}

We will look for a time dependence of the cosmological scale factor 
$a(t)$ (which determines the determinant of the spatial metric  
$\gamma_{\alpha\beta}$)  near the singularity in the following form:
\begin{equation}
a(t) = \sum_{n=0} a_n t^{\gamma_n}.
\label{radius}
\end{equation}
Substituting this expansion into the Friedmann equation 
(\ref{Friedmann}) and comparing the smallest (i.e., most singular) 
powers in $t$, one can easily find that 
\begin{equation}
\gamma_0 =\frac{2}{\alpha_2}.
\label{gamma0}
\end{equation}
Thus, the lowest exponent in the quasi-isotropic solution is defined by 
the stiffer fluid. The next term in the expansion (\ref{radius}) 
arises due to the presence of the second fluid in the Universe. 
Its time dependence is characterized by the exponent 
\begin{equation}
\gamma_1 = 2 + \gamma_0 (1-\alpha_1) = 2 + \frac{2}{\alpha_2} - 
\frac{2 \alpha_1}{\alpha_2}.
\label{gamma1}
\end{equation}
The third term in the expansion (\ref{radius}) arises due to the 
presence of the curvature term. It has the same exponent as in the 
case of the one - fluid cosmological model \cite{we}:
\begin{equation}
\gamma_2 = 2 - \gamma_0 = 2 - \frac{2}{\alpha_2}.
\label{gamma2}
\end{equation}
The next term of the expansion is again connected with the presence 
of the second fluid. The correspondent exponent is equal to 
\begin{equation}
\gamma_3 = 4 + \gamma_0 - 2\gamma_0 \alpha_1 = 4 + \frac{2}
{\alpha_2} - \frac{4 \alpha_1}{\alpha_2}.
\label{gamma3}
\end{equation}
Notice that three cases are possible depending on relation between
$\alpha_1$ and $\alpha_2$:
\begin{equation}
1)~~~\gamma_3 > \gamma_2,~~~\alpha_2 >  2\alpha_1 - 2~, 
\label{case1}
\end{equation}
e.g., if $k_2 = 1,~\alpha_2 = 6$ and $k_1 = 0,~\alpha_1 = 3$, i.e., 
the dominant fluid is ultra-stiff matter and the subdominant fluid is 
dust;
\begin{equation}
2)~~~\gamma_3 = \gamma_2,~~~\alpha_2 = 2\alpha_1 - 2~,  
\label{case2}
\end{equation}
e.g., in the case $k_2 = 1,~\alpha_2 = 6$ and $k_1 = 1/3,~ 
\alpha_1 = 4$, so the first fluid is ultra-stiff matter and the second 
fluid is radiation, or in the dust--radiation case  
$k_2 = 1/3,~\alpha_2 = 4$ and $k_1 = 0,~\alpha_1 = 3$;
\begin{equation}
3)~~~\gamma_3 < \gamma_2,~~~\alpha_2 < 2\alpha_1 - 2~,  
\label{case3}
\end{equation}
(e.g., in the case $k_2 = 1,~\alpha_2 = 6$ and $k_1 = 2/3,~
\alpha_1 = 5$).

Now it is easy to understand that all terms which are present in the 
series expansion for the one-fluid quasi-isotropic solution remain in 
the solution for a two-fluid case because the mechanism of their 
appearance remains the same: interference between the dominant fluid 
and the curvature term. It is convenient to write a general formula 
for this set of terms in the following form:
\begin{equation}
\gamma_{n,0} = \gamma_0 + n(2 - 2\gamma_0), \ n=0,1,\ldots.
\label{gamman0}
\end{equation}
Another sequence of terms in the quasi-isotropic solution arises 
due to interference between the two fluids. It has the following set 
of exponents: 
\begin{equation}
\gamma_{0,m} = \gamma_0 + m(2 - \gamma_0 \alpha_1), \ n=1,\ldots.
\label{gamma0m}
\end{equation}
In addition, the full set of terms occurring in the quasi-isotropic 
solution includes terms arising from mixing between the two sequences 
(\ref{gamman0}) and (\ref{gamma0m}).  Thus, the expansion  
(\ref{radius}) can be represented as 
\begin{equation}
a(t) = \sum_{n,m=0} a_{n,m} t^{\gamma_{n,m}}
\label{radius1}
\end{equation}
where 
\begin{equation}
\gamma_{n,m} = \gamma_0 + n(2-2\gamma_0) + m(2-\gamma_0 \alpha_1).
\label{gammanm}
\end{equation}
It is easy to see that if the number 
\begin{equation}
\frac{2-2\gamma_0}{2-\gamma_0 \alpha_1} = \frac{\alpha_2 - 2}
{\alpha_2 - \alpha_1} = \frac{3k_2 + 1}{3(k_2-k_1)} 
\label{degeneration}
\end{equation}
is rational, a degeneracy between exponents $\gamma_{n,m}$ in 
Eq.~(\ref{gammanm}) is possible (an example of such degeneracy is 
given in Eq. (\ref{case2})). However, it does not create any 
peculiarities in the structure of the quasi-isotropic solution. It 
is important that the expansion (\ref{radius1}) with the exponents 
given in Eq.~(\ref{gammanm}) contains all possible cross-terms. 

Correspondingly, the inhomogeneous quasi-isotropic solution for 
the spatial metric $\gamma_{\alpha\beta}$ can be represented as 
\begin{equation}
\gamma_{\alpha\beta} = \sum_{n,m =0} \gamma_{\alpha\beta}^{(n,m)} 
t^{2\gamma_0 + n(2-2\gamma_0) + m(2-\gamma_0 \alpha_1)}.
\label{quasi}
\end{equation}

 For illustrative purposes, let us display the structure of this 
solution for some simple two-fluid models. For the model with 
radiation and dust, the series (\ref{quasi}) can be rewritten as 
\begin{equation}
\gamma_{\alpha\beta} = \sum_{n =0} \gamma_{\alpha\beta}^{(n)} 
t^{1+n/2}~.
\label{quasi1}
\end{equation}
 For ultra-stiff matter and radiation, one has
\begin{equation}
\gamma_{\alpha\beta} = \sum_{n =0} \gamma_{\alpha\beta}^{(n)} 
t^{2/3+2n/3}~.
\label{quasi2}
\end{equation}
In the case of ultra-stiff matter and dust, we get
\begin{equation}
\gamma_{\alpha\beta} = \gamma_{\alpha\beta}^{(0)}t^{2/3} 
+ \gamma_{\alpha\beta}^{(1)}t^{5/3} + \gamma_{\alpha\beta}^{(2)}
t^{2} + \sum_{n =3} \gamma_{\alpha\beta}^{(n)} t^{8/3+(n-3)/3}~,
\label{quasi3}
\end{equation}
and for stiff matter and the fluid with $k_1 = 2/3$, the solution reads
\begin{equation}
\gamma_{\alpha\beta} = \sum_{n =0} \gamma_{\alpha\beta}^{(n)} 
t^{2/3+n/3}~.
\label{quasi4}
\end{equation}

Now, we will deduce explicit formulae for the first terms of the 
quasi-isotropic solution. The following notations will be used:
\begin{equation}
\gamma_{\alpha\beta} = a_{\alpha\beta}t^{2\gamma_0} +
b_{\alpha\beta} t^{2\gamma_0 + 2-\gamma_0 \alpha_1}
+ c_{\alpha\beta}t^2
+ d_{\alpha\beta}t^{2\gamma_0 + 4-2\gamma_0 \alpha_1} + \cdots.
\label{quasi5}
\end{equation}
As usually, the spatial metric $a_{\alpha\beta}$ is arbitrary. For the 
time being we treat the metric $b_{\alpha\beta}$ in the second term of 
the expansion (\ref{quasi5}) arising due to the presence of the second 
fluid as an arbitrary positively defined symmetric tensor.

The inverse spatial metric reads
\begin{eqnarray}
&&\gamma^{\alpha\beta} = a^{\alpha\beta} t^{-2\gamma_0} - 
b^{\alpha\beta}t^{2-\gamma_0\alpha_1-2\gamma_0}\nonumber \\
&&-c^{\alpha\beta}t^{2-4\gamma_0} 
-d^{\alpha\beta}t^{4-2\gamma_0\alpha_1-2\gamma_0}
+b^{\alpha}_{\delta}b^{\delta\beta}t^{4-2\gamma_0\alpha_1-2\gamma_0}
\label{inverse}
\end{eqnarray}
where $a^{\alpha\beta}$ is defined by the relation
\begin{equation}
a^{\alpha\beta} a_{\beta\gamma} = \delta_{\gamma}^{\alpha},
\label{inverse1}
\end{equation}
while the indices of all other matrices are lowered and raised by
$a_{\alpha\beta}$ and $a^{\alpha\beta}$, for example,
\begin{equation}
b_{\beta}^{\alpha} = a^{\alpha\gamma}b_{\gamma\beta}.
\end{equation}
Let us also write down the expressions for the extrinsic curvature, 
its contractions and its derivatives:
\begin{eqnarray}
&&\kappa_{\alpha\beta} \equiv \frac{\partial \gamma_{\alpha\beta}}
{\partial t}= 2\gamma_0 a_{\alpha\beta}t^{2\gamma_0-1} + (2\gamma_0 + 
2-\gamma_0 \alpha_1) b_{\alpha\beta} t^{2\gamma_0+1-\gamma_0\alpha_1}
\nonumber \\
&&+2c_{\alpha\beta} t + (2\gamma_0 + 4 - 2\gamma_0\alpha_1) 
d_{\alpha\beta}t^{2\gamma_0 + 3 - 2\gamma_0\alpha_1}, 
\label{extrinsic}
\end{eqnarray}
\begin{eqnarray}
&&\kappa_{\alpha}^{\beta} = 2\gamma_0 \delta_{\alpha}^{\beta} t^{-1} 
+ (2-\gamma_0\alpha_1)b_{\alpha}^{\beta}t^{1-\gamma_0\alpha_1}
+2(1-\gamma_0)c_{\alpha}^{\beta}t^{1-2\gamma_0}\nonumber \\
&&+2(2-\gamma_0\alpha_1)d_{\alpha}^{\beta}t^{3-2\gamma_0\alpha_1} - 
(2 - \gamma_0\alpha_1) b_{\alpha\delta}b^{\delta\beta}t^{3-
2\gamma_0\alpha_1},
\label{extrinsic1}
\end{eqnarray}
\begin{eqnarray}
&&\kappa_{\alpha}^{\alpha} = 6\gamma_0 t^{-1} + 
(2-\gamma_0\alpha_1)bt^{1-\gamma_0\alpha_1}
+2(1-\gamma_0)c_{\alpha}^{\beta}t^{1-2\gamma_0}\nonumber \\
&&+2(2-\gamma_0\alpha_1)d t^{3-2\gamma_0\alpha_1} - (2 - 
\gamma_0\alpha_1) b_{\alpha\beta}b^{\alpha\beta}t^{3-2\gamma_0
\alpha_1},
\label{extrinsic2}
\end{eqnarray}
\begin{eqnarray}
&&\frac{\partial\kappa_{\alpha}^{\beta}}{\partial t} =
-2\gamma_0\delta_{\alpha}^{\beta}t^{-2} + (2-\gamma_0\alpha_1)
(1-\gamma_0\alpha_1)b_{\alpha}^{\beta}t^{-\gamma_0\alpha_1}\nonumber \\
&&+ 2(1-\gamma_0)(1-2\gamma_0)c_{\alpha}^{\beta}t^{-2\gamma_0}
+2(2-\gamma_0\alpha_1)(3-2\gamma_0\alpha_1)d_{\alpha}^{\beta}
t^{2-2\gamma_0\alpha_1}\nonumber \\
&&-(2-\gamma_0\alpha_1)(3-2\gamma_0\alpha_1)b_{\alpha\delta}
b^{\delta\beta}t^{2-2\gamma_0\alpha_1},
\label{extrinsic3}
\end{eqnarray}
\begin{eqnarray}
&&\frac{\partial\kappa_{\alpha}^{\alpha}}{\partial t} =
-6\gamma_0t^{-2}+ (2-\gamma_0\alpha_1)(1-\gamma_0\alpha_1)b
t^{-\gamma_0\alpha_1}
\nonumber \\
&&+ 2(1-\gamma_0)(1-2\gamma_0)ct^{-2\gamma_0}
+2(2-\gamma_0\alpha_1)(3-2\gamma_0\alpha_1)dt^{2-2\gamma_0
\alpha_1}\nonumber \\
&&-(2-\gamma_0\alpha_1)(3-2\gamma_0\alpha_1)b_{\alpha\delta}
b^{\delta\alpha}t^{2-2\gamma_0\alpha_1},
\label{extrinsic4}
\end{eqnarray}
\begin{eqnarray}
&&\kappa_{\alpha}^{\beta}\kappa_{\beta}^{\alpha} = 12\gamma_0^2
t^{-2} + 4\gamma_0 (2-\gamma_0\alpha_1)bt^{-\gamma_0
\alpha_1} \nonumber \\
&&+8\gamma_0(1-\gamma_0)ct^{-2\gamma_0}
+8\gamma_0(2-\gamma_0\alpha_1)dt^{2-2\gamma_0\alpha_1}\nonumber \\
&&+(2-\gamma_0\alpha_1)(2-\gamma_0\alpha_1-4\gamma_0)b_{\alpha}^{\beta}
b_{\beta}^{\alpha}t^{2-2\gamma_0\alpha_1}.
\label{extrinsic5}
\end{eqnarray}
Also, we need an explicit expression for the determinant of the spatial
metric and its time derivative:
\begin{eqnarray}
&&\gamma \equiv \det \gamma_{\alpha\beta} = t^{6\gamma_0}\det a 
(1+bt^{2-\gamma_0\alpha_1} + ct^{2-2\gamma_0} \nonumber \\
&&+d t^{4-2\gamma_0\alpha_1} + (1/2)(b^2- b_{\alpha}^{\beta}
b_{\beta}^{\alpha})t^{4-2\gamma_0\alpha_1}),
\label{determin}
\end{eqnarray}
\begin{eqnarray}
&&\frac{\dot{\gamma}}{\gamma} = 6\gamma_0 t^{-1} + (2-\gamma_0\alpha_1)
bt^{1-\gamma_0\alpha_1} +2(1-\gamma_0)ct^{1-2\gamma_0}\nonumber \\
&&+2(2-\gamma_0\alpha_1)dt^{3-2\gamma_0\alpha_1}- (2-\gamma_0\alpha_1)
b_{\alpha}^{\beta}b_{\beta}^{\alpha}t^{3-2\gamma_0\alpha_1}.
\label{determin2}
\end{eqnarray}

Now, using well-known expressions for the components of the Ricci tensor
\cite{Land-Lif}:
\begin{equation}
R_0^0 = -\frac{1}{2} \frac{\partial \kappa_{\alpha}^{\alpha}}
{\partial t} - \frac{1}{4}\kappa_{\alpha}^{\beta}\kappa_{\beta}^{\alpha},
\label{curvature}
\end{equation}
\begin{equation}
R_{\alpha}^0 = \frac{1}{2}(\kappa_{\alpha;\beta}^{\beta} -
\kappa_{\beta;\alpha}^{\beta}),
\label{curvature1}
\end{equation}
\begin{equation}
R_{\alpha}^{\beta} = -P_{\alpha}^{\beta} -\frac{1}{2}\frac{\partial
\kappa_{\alpha}^{\beta}}{\partial t} - \frac{\dot{\gamma}}{4\gamma}
\kappa_{\alpha}^{\beta}
\label{curvature2}
\end{equation}
where $P_{\alpha}^{\beta}$ is the three-dimensional part of the Ricci 
tensor, and substituting the expressions (\ref{extrinsic})-
(\ref{determin2}) into Eqs. (\ref{curvature})-(\ref{curvature2}), one 
immediately obtains
\begin{eqnarray}
&&R_0^0 = 3\gamma_0(1-\gamma_0)t^{-2} -\frac{1}{2}(2-\gamma_0\alpha_1)
(1-\gamma_0\alpha_1+2\gamma_0)bt^{-\gamma_0\alpha_1} \nonumber \\
&&-(1-\gamma_0)ct^{-2\gamma_0}-(2-\gamma_0\alpha_1)(3-2\gamma_0\alpha_1
+2\gamma_0)dt^{2-2\gamma_0\alpha_1}
\nonumber \\
&&+\frac{1}{4}(2-\gamma_0\alpha_1) (4-3\gamma_0\alpha_1+4\gamma_0)
b_{\alpha}^{\beta}b_{\beta}^{\alpha}t^{2-2\gamma_0\alpha_1}, 
\label{curvature3}
\end{eqnarray}
\begin{eqnarray}
&&R_{\alpha}^0 = \frac12(2-\gamma_0\alpha_1)(b^{\beta}_{\alpha;\beta}
-b_{;\alpha})t^{1-\gamma_0\alpha_1} \nonumber \\
&&+(1-\gamma_0)(c^{\beta}_{\alpha;\beta}-c_{;\alpha})t^{1-2\gamma_0}
+(2-\gamma_0\alpha_1) (d^{\beta}_{\alpha;\beta}-d_{;\alpha})
t^{3-2\gamma_0\alpha_1}\nonumber\\ 
&&+ \frac12(2-\gamma_0\alpha_1)((b_{\mu}^{\nu}b_{\nu}^{\mu})_{;\alpha}
-(b_{\alpha}^{\mu}b_{\mu}^{\beta})_{;\beta})t^{3-2\gamma_0\alpha_1}.
\label{curvature4}
\end{eqnarray}
To write down an explicit expression for $R_{\alpha}^{\beta}$, we 
need an expression for $P_{\alpha}^{\beta}$ taken up to linear 
terms in $b_{\alpha\beta}$. It reads
\begin{eqnarray}
&&P_{\alpha}^{\beta} = \tilde{P}_{\alpha}^{\beta}t^{-2\gamma_0} - 
b^{\mu\nu}\tilde{P}^{\beta}_{\ \mu\alpha\nu}t^{2-\gamma_0\alpha_1-
2\gamma_0}
\nonumber \\
&&+\frac12(2b^{\nu\beta}_{;\nu\alpha}-b^{\ \beta}_{;\ \alpha}
-b^{\nu\beta}_{;\alpha\nu} - b_{\alpha;\nu}^{\beta\ \nu}
+b_{\alpha;\ \nu}^{\nu\ \beta})t^{2-\gamma_0\alpha_1-2\gamma_0}.
\label{tensorP}
\end{eqnarray} 
Here, $\tilde{P}^{\beta}_{\ \mu\alpha\nu}$ denotes the three-dimensional 
Riemann-Christoffel tensor for the spatial metric $a_{\alpha\beta}$. 
Covariant derivatives are also taken with respect to this metric. 
Now
\begin{eqnarray}
&&R_{\alpha}^{\beta} = -\tilde{P}_{\alpha}^{\beta}t^{-2\gamma_0} 
+ b^{\mu\nu}\tilde{P}^{\beta}_{\ \mu\alpha\nu}t^{2-\gamma_0\alpha_1
-2\gamma_0}
\nonumber \\
&&-\frac12(2b^{\nu\beta}_{;\nu\alpha}-b^{\ \beta}_{;\ \alpha}
-b^{\nu\beta}_{;\alpha\nu} - b_{\alpha;\nu}^{\beta\ \nu}
+b_{\alpha;\ \nu}^{\nu\ \beta})t^{2-\gamma_0\alpha_1-2\gamma_0}
\nonumber\\
&&+\gamma_0(1-3\gamma_0)\delta_{\alpha}^{\beta}t^{-2}
-\frac12(2-\gamma_0\alpha_1)((1-\gamma_0\alpha_1+3\gamma_0)
b_{\alpha}^{\beta}
+\gamma_0b\delta_{\alpha}^{\beta})t^{-\gamma_0\alpha_1}\nonumber \\
&&-(1-\gamma_0)((1-\gamma_0)c_{\alpha}^{\beta}+\gamma_0
c\delta_{\alpha}^{\beta})
t^{-2\gamma_0}\nonumber \\
&&-(2-\gamma_0\alpha_1)((3-2\gamma_0\alpha_1+3\gamma_0)
d_{\alpha}^{\beta}+\gamma_0d\delta_{\alpha}^{\beta})t^{2-2\gamma_0
\alpha_1}\nonumber \\
&&+\frac12(2-\gamma_0\alpha_1)\left((3-2\gamma_0\alpha_1+3\gamma_0)
b_{\alpha}^{\nu}b_{\nu}^{\beta} +\gamma_0b_{\mu}^{\nu}b_{\nu}^{\mu}
\delta_{\alpha}^{\beta}\right.\nonumber \\
&&\left.-\frac12(2-\gamma_0\alpha_1)bb_{\alpha}^{\beta}\right)
t^{2-2\gamma_0\alpha_1}.
\label{curvature5}
\end{eqnarray}

The Einstein equations are, as usually,  
\begin{equation}
R_i^j = 8\pi G (T_i^j - \frac{1}{2}\delta_i^j T)
\label{Einstein}
\end{equation}
where the energy-momentum tensor for a two-fluid model has the form
\begin{equation}
T_{i}^{j} = (\varepsilon_1 + p_1)u_{1i} u^{1j} - p_1 \delta_{i}^{j}
+(\varepsilon_2 + p_2)u_{2i} u^{2j} - p_2 \delta_{i}^{j}.
\label{tensor}
\end{equation}
Analyzing the Friedmann equation (\ref{Friedmann}) again, one can see 
that series expansions for the energy densities (and, hence, for the 
pressures) have the following form: 
\begin{equation}
\sum_{n,m=0}\varepsilon_1^{(n,m)}(x) t^{-\gamma_0\alpha_1+n(2-2\gamma_0)
+m(2-\gamma_0\alpha_1)},
\label{energy1}
\end{equation} 
\begin{equation}
\sum_{n,m=0}\varepsilon_2^{(n,m)}(x) t^{-2+n(2-2\gamma_0)
+m(2-\gamma_0\alpha_1)}.
\label{energy2}
\end{equation} 
It is easy to notice that the series expansion (\ref{energy2}) contains 
all powers of $t$ which appear in (\ref{energy1}). However, the series 
(\ref{energy2}) contains the term proportional to $1/t^2$ which is 
absent in the series for $\varepsilon_1$. Besides, the series 
(\ref{energy2}) contains the terms $\varepsilon_2^{(n,0)}
t^{-2+n(2-2\gamma_0)},~n=1,2,\ldots$ which could appear in the series 
(\ref{energy1}) in the case of the degeneracy of the type given by 
Eq.~(\ref{degeneration}).

Now let us write down components of the Einstein equations 
(\ref{Einstein}) in more detail:
\begin{eqnarray}
&&R_0^0 = 4\pi G((1+3k_1)\varepsilon_1 - 2(1+k_1)\varepsilon_1
u_{1\alpha}u_1^{\alpha}\nonumber \\
&&+(1+3k_2)\varepsilon_2 - 2(1+k_2)\varepsilon_2u_{1\alpha}u_2^{\alpha}),
\label{Einstein1}
\end{eqnarray}
\begin{equation}
R_{\alpha}^0 = 8\pi G((1+k_1)\varepsilon_1 u_{1\alpha}u_1^0 +  
(1+k_2)\varepsilon_2 u_{2\alpha}u_2^0),
\label{Einstein2}
\end{equation}
\begin{eqnarray}
&&R_{\alpha}^{\beta} = 4\pi G (2(1+k_1)\varepsilon_1
u_{1\alpha}u_1^{\beta} - 
(1-k_1)\varepsilon_1 \delta_{\alpha}^{\beta}\nonumber \\
&&+ 2(1+k_2)\varepsilon_2u_{2\alpha}u_2^{\beta} - 
(1-k_2)\varepsilon_2 \delta_{\alpha}^{\beta}).
\label{Einstein3}
\end{eqnarray}  

Comparing the structure of the left-hand side of Eq.~(\ref{Einstein2}) 
with its right-hand side, one can see that the series structure for 
spatial components of fluid four-velocities $u_{\alpha}$ has the 
following form:
\begin{equation}
u_{2\alpha} = \sum_{n,m=0}u_{2\alpha}^{(n,m)}t^{1+n(2-\gamma_0)
+m(2-\gamma_0\alpha_1)}
\label{velocity}
\end{equation}
where $u_{2\alpha}^{(0,0)} = 0$,
and 
\begin{equation}
u_{1\alpha} = \sum_{n,m=0}u_{1\alpha}^{(n,m)}t^{1+n(2-\gamma_0)
+m(2-\gamma_0\alpha_1)}.
\label{velocity1}
\end{equation}
Correspondingly, 
\begin{equation}
u_2^{\alpha} = \sum_{n,m=0}u_2^{\alpha(n,m)}t^{1-2\gamma_0
+n(2-\gamma_0)+m(2-\gamma_0\alpha_1)}
\label{velocity2}
\end{equation}
where $u_2^{\alpha(0,0)} = 0$,
and 
\begin{equation}
u_1^{\alpha} = \sum_{n,m=0}u_1^{\alpha(n,m)}t^{1-2\gamma_0+n(2-\gamma_0)
+m(2-\gamma_0\alpha_1)}.
\label{velocity3}
\end{equation}
Also, we need the following expressions:
\begin{equation}
u_{20} = u_2^0 = 1 + \sum_{n,m=0}u_{20}^{(n,m)} t^{2-2\gamma_0 
+n(2-\gamma_0)+m(2-\gamma_0\alpha_1)}
\label{velocity4}
\end{equation}
where $u_{20}^{(0,0)} = 0$, 
and 
\begin{equation}
u_{10} = u_1^0 = 1 + \sum_{n,m=0}u_{10}^{(n,m)} t^{2-2\gamma_0 
+n(2-\gamma_0)+m(2-\gamma_0\alpha_1)}.
\label{velocity5}
\end{equation}
Now, taking the leading term proportional to $t^{-2}$ in the expansion 
for $R_0^0$ and comparing it with the corresponding terms in the series 
for the energy-momentum tensor, one can find from Eq.~(\ref{Einstein1}) 
that the leading term in the series for the density energy (\ref{energy2}) 
is given by the formula 
\begin{equation}
\varepsilon_2^{(0,0)} = \frac{1}{6\pi G(1+k_2)^2}.
\label{energy200}
\end{equation}
Thus, the firstly dominant (stiffer) component becomes homogeneous
at $t\to 0$. 

The next step consists in comparison of the terms proportional to 
$t^{-2\gamma_0\alpha_1}$ in the equations (\ref{Einstein1}) and 
(\ref{Einstein3}). As a result, two consistency relations for the two 
unknown variables $\varepsilon_2^{(0,1)}$ and $\varepsilon_1^{(0,0)}$ 
arise:
\begin{eqnarray}
&&-(2-\gamma_0\alpha_1)(1-\gamma_0\alpha_1+2\gamma_0)b \nonumber\\
&&= 8\pi G((1+3k_1)\varepsilon_1^{(0,0)}+(1+3k_2)\varepsilon_2^{(0,0)},
\label{eq1}
\end{eqnarray}
\begin{eqnarray} 
&&(2-\gamma_0\alpha_1)((1-\gamma_0\alpha_1+3\gamma_0)b_{\alpha}^{\beta}
+\gamma_0b\delta_{\alpha}^{\beta}) \nonumber \\
&&= 8\pi G( (1-k_1)\varepsilon_1^{(0,0)} \delta_{\alpha}^{\beta}+
(1-k_2)\varepsilon_2^{(0,1)} \delta_{\alpha}^{\beta}).
\label{eq2}
\end{eqnarray}
It follows from Eq. (\ref{eq2}) that 
\begin{equation}
b_{\alpha}^{\beta} = \frac13 b \delta_{\alpha}^{\beta}.
\label{restrictionb}
\end{equation}
This means that we cannot choose the metric $b_{\alpha\beta}$ freely;
it should be proportional to the metric $a_{\alpha\beta}$, i.e.,
\begin{equation}
b_{\alpha\beta} = \frac13 b a_{\alpha\beta}.
\label{restrictionb1}
\end{equation}
Thus, only one new arbitrary function of spatial coordinates $b(\bf r)$
appears in the second term of the series in the right-hand side of
Eq.~(\ref{quasi5}). We will immediately see that it describes the
amplitude of {\em isocurvature} perturbations. Really, substituting 
the expression (\ref{restrictionb1}) into Eq. (\ref{eq2}), one get
\begin{eqnarray} 
&&(2-\gamma_0\alpha_1)(1-\gamma_0\alpha_1+6\gamma_0)b \nonumber \\
&&= 24\pi G( (1-k_1)\varepsilon_1^{(0,0)}+
(1-k_2)\varepsilon_2^{(0,1)})
\label{eq21}
\end{eqnarray}
and then, resolving Eqs. (\ref{eq1}) and (\ref{eq21}) with respect to 
$\varepsilon_1^{(0,0)}$ and $\varepsilon_2^{(0,1)}$, we obtain
\begin{equation}
\varepsilon_1^{(0,0)} = \frac{(3k_2-2k_1+1)b}{12\pi G(k_2+1)^2},
\label{energy100}
\end{equation}
\begin{equation}
\varepsilon_2^{(0,1)} = - \frac{b}{12\pi G(k_2+1)}.
\label{energy201}
\end{equation}
Therefore, the firstly subdominant component 1 is generically 
essentially inhomogeneous near the singularity.

Now, we should compare terms proportional to $t^{-2\gamma_0}$ in the 
left- and right-hand sides of Eqs.~(\ref{Einstein1}) and 
(\ref{Einstein3}). We consider a generic situation when the degeneracy 
described above (see Eq. (\ref{case2})) is absent. Then we have 
\begin{equation}
-(1-\gamma_0)c = 4\pi G(1+3k_2)\varepsilon_2^{(1,0)},
\label{eq3}
\end{equation}
\begin{eqnarray}
&&\tilde{P}_{\alpha}^{\beta}+(1-\gamma_0)((1+\gamma_0)c_{\alpha}^{\beta}
+\gamma_0c\delta_{\alpha}^{\beta})
\nonumber \\
&&=4\pi G (1-k_2)\varepsilon_2^{(1,0)} \delta_{\alpha}^{\beta}.
\label{eq4}
\end{eqnarray}
Resolving Eqs. (\ref{eq3}), (\ref{eq4}) with respect to variables 
$c_{\alpha}^{\beta}$ and $\varepsilon_2^{(1,0)}$, one obtains the 
following expressions which coincide with known expressions for the 
one-fluid quasi-isotropic solution \cite{we}:
\begin{equation}
c_{\alpha}^{\beta} = -\frac{9(k_2+1)^2}{(3k_2+1)(3k_2+5)}
\left(\tilde{P}_{\alpha}^{\beta} +\frac{\tilde{P}
(3k_2^2-6k_2-5)\delta_{\alpha}^{\beta}}{4(9k_2+5)}\right),
\label{c}
\end{equation}
\begin{equation}
\varepsilon_2^{(1,0)} = \frac{3\tilde{P}(k_2+1)}{16\pi G(9k_2+5)}.
\label{energy210}
\end{equation}

The next step consists in the consideration of terms proportional to 
$t^{2-2\gamma_0\alpha_1}$. Using Eq. (\ref{restrictionb}), one can 
obtain from Eqs.~(\ref{Einstein1}) and (\ref{Einstein3}) that 
\begin{eqnarray}
&&-(2-\gamma_0\alpha_1)(3-2\gamma_0\alpha_1+2\gamma_0)d
+ \frac{(2-\gamma_0\alpha_1)(4-3\gamma_0\alpha_1+4\gamma_0)b^2}
{12}\nonumber \\
&&=4\pi G(\varepsilon_2^{(0,2)}(1+3k_2) + \varepsilon_1^{(0,1)}(1+3k_1)),
\label{eq5}
\end{eqnarray}
\begin{eqnarray}
&&(2-\gamma_0\alpha_1)((3-2\gamma_0\alpha_1+3\gamma_0)d_{\alpha}^{\beta} 
+\gamma_0 d\delta_{\alpha}^{\beta})\nonumber \\
&&+\frac{(2-\gamma_0\alpha_1)(12-7\gamma_0\alpha_1+12\gamma_0)
b^2\delta_{\alpha}^{\beta}}{36}
\nonumber \\ 
&&=4\pi G\delta_{\alpha}^{\beta}(\varepsilon_1^{(0,1)}(1-k_1)
+ \varepsilon_2^{(0,2)}(1-k_2)).
\label{eq6}
\end{eqnarray}
It is easy to see from Eq. (\ref{eq6}) that
\begin{equation}
d_{\alpha}^{\beta} = \frac13 d\delta_{\alpha}^{\beta}.
\label{restrictiond}
\end{equation}
Substituting Eq. (\ref{restrictiond}) into Eq. (\ref{eq6}), one get 
\begin{eqnarray}
&&\frac{(2-\gamma_0\alpha_1)(3-2\gamma_0\alpha_1+6\gamma_0)
d\delta_{\alpha}^{\beta}}{3}\nonumber \\
&&+\frac{(2-\gamma_0\alpha_1)(12-7\gamma_0\alpha_1+12\gamma_0)b^2
\delta_{\alpha}^{\beta}}{36}
\nonumber \\ 
&&=4\pi G\delta_{\alpha}^{\beta}(\varepsilon_1^{(0,1)}(1-k_1)
+ \varepsilon_2^{(0,2)}(1-k_2)).
\label{eq61}
\end{eqnarray}

Now, using the definitions of the coefficients $\alpha_1,\alpha_2$ and 
$\gamma_0$, it is convenient to rewrite Eqs. (\ref{eq5}) and (\ref{eq61}) 
in the following form:
\begin{eqnarray}
&&\varepsilon_2^{(0,2)}(1+3k_2) + \varepsilon_1^{(0,1)}(1+3k_1)\nonumber \\
&&=-\frac{(k_2-k_1)(9k_2-12k_1+1)d}{6\pi G(k_2+1)^2} 
+\frac{(k_2-k_1)(6k_2-9k_1+1)}{36\pi G(k_2+1)^2},
\label{eq52}
\end{eqnarray}
\begin{eqnarray} 
&&\varepsilon_2^{(0,2)}(1-k_2) + \varepsilon_1^{(0,1)}(1-k_1) \nonumber \\
&&=\frac{(k_2-k_1)(3k_2-4k_1+3)d}{6\pi G(k_2+1)^2}
+\frac{(k_2-k_1)(k_1-3)}{36\pi G(k_2+1)^2}.
\label{eq62}
\end{eqnarray}

The system of equations (\ref{eq52}) and (\ref{eq62}) contains three 
variables $d,\varepsilon_2^{(0,2)}$ and $\varepsilon_1^{(0,1)}$.
To solve it we use the energy conservation law
\begin{equation}
T_{0\ ;i}^i = 0.
\label{conserv}
\end{equation}
In the case of non-interacting perfect fluids, Eq. (\ref{conserv}) gives 
two separate equations in our approximation:
\begin{equation}
\dot{\varepsilon}_2 + \frac{1}{2}\frac{\dot{\gamma}}{\gamma}(\varepsilon_2 
+ p_2) = 0;
\label{conserv1}
\end{equation}
\begin{equation}
\dot{\varepsilon}_1 + \frac{1}{2}\frac{\dot{\gamma}}{\gamma}(\varepsilon_1 
+ p_1) = 0.
\label{conserv2}
\end{equation}
Using the series (\ref{energy1}), (\ref{energy2}) and (\ref{determin2}), 
we get the following formulae connecting $\varepsilon_2^{(0,2)},
\varepsilon_1^{(0,1)},d$ and $b$:
\begin{equation}
\varepsilon_2^{(0,2)} = -\frac{d}{12\pi G(1+k_2)} + \frac{(3k_2+5)b^2}
{144\pi G(1+k_2)},
\label{energy2d}
\end{equation}
\begin{equation}
\varepsilon_1^{(0,1)} = -\frac{(k_1+1)(3k_2-2k_1+1)b^2}{24\pi G(1+k_2)^2}.
\label{energy101}
\end{equation}
Substituting expressions for the energy densities from Eqs.~(\ref{energy2d}) 
and (\ref{energy101}) into Eq.~(\ref{eq52}), one get the following 
expression for $d$:
\begin{equation}
d = \frac{(-3k_2^2+6k_1^2-6k_1k_2-2k_2-2k_1-1)b^2}{12(5k_2-4k_1+1)}.
\label{d}
\end{equation}
It is easy to check that this expression satisfies Eq.~(\ref{eq62}), too.
Substituting (\ref{d}) into Eq. (\ref{energy2d}), we have
\begin{equation}
\varepsilon_2^{(0,2)} = \frac{(3k_2^2-k_1^2-k_1k_2+5k_2-3k_1+1)b^2}
{24\pi G(1+k_2)(5k_2-4k_1+1)}.
\label{energy202}
\end{equation}  

Now, we would like to write down explicit formulae for first coefficients 
of the velocity series (\ref{velocity}), (\ref{velocity1}). Using these 
series and the series for the energy densities  (\ref{energy1}), 
(\ref{energy2}), one can get the following relation for terms proportional 
to $t^{1-\gamma_0\alpha_1}$ in the Einstein equation (\ref{Einstein2}):
\begin{equation}
\frac12(2-\gamma_0\alpha_1)(b_{\alpha\ ;\beta}^{\beta} - b_{;\alpha}) =
8\pi G((1+k_2)\varepsilon_2^{(0,0)}u_{2\alpha}^{(0,1)} + 
(1+k_1)\varepsilon_{1\alpha}^{(0,0)}u_{1\alpha}^{(0,0)}).
\label{eq-vel}
\end{equation}
Eq. (\ref{eq-vel}) contains two unknown quantities $u_{2\alpha}^{(0,0)}$ 
and $u_{1\alpha}^{(0,0)}$. Thus, we need additional relations to resolve 
it. We will use remaining components of the energy-momentum conservation 
law:
\begin{equation}
T_{\alpha\ ; i}^{i} = 0.
\label{Bianchi}
\end{equation}
In the synchronous system of reference, this equation can be rewritten as 
\begin{equation}
T_{\alpha\ ,0}^{0} + T_{\alpha\ ;\beta}^{\beta} + \frac12 k_{\beta}^{\beta} 
T_{\alpha}^{0} = 0.
\label{Bianchi1}
\end{equation}

Using the definition of energy-momentum tensor components for 
perfect fluids and the explicit expression for the trace of the extrinsic 
curvature (\ref{extrinsic2}), we have the following couple of relations in 
the needed order of perturbation theory:
\begin{equation}
k_2 \varepsilon_{2\ \ ;\alpha}^{(0,1)} = (1+k_2)\frac{\partial}{\partial t}
 (\varepsilon_2^{(0,0)}u_{2\alpha}^{(0,1)}) + \frac{3\gamma_0}{t}(1 + k_2)
 \varepsilon_2^{(0,0)}u_{2\alpha}^{(0,1)},
\label{eq-vel1}  
\end{equation}
\begin{equation}
k_1 \varepsilon_{1\ \ ;\alpha}^{(0,0)} = (1+k_1)\frac{\partial}{\partial t}
 (\varepsilon_1^{(0,0)}u_{1\alpha}^{(0,0)}) + \frac{3\gamma_0}{t}(1 + k_2)
 \varepsilon_1^{(0,0)} u_{1\alpha}^{(0,0)}.
\label{eq-vel2}  
\end{equation}
Substituting the expressions (\ref{energy200}), (\ref{energy201}) and 
(\ref{energy100}) for the coefficients $\varepsilon_2^{(0,0)},\varepsilon_2^{(0,1)}$ and $\varepsilon_1^{(0,0)}$
into Eqs. (\ref{eq-vel1}) and (\ref{eq-vel2}), one finds
\begin{equation}
u_{2\alpha}^{(0,1)} = -\frac{k_2(k_2+1)b_{;\alpha}}{2(k_2-2k_1+1)},
\label{velocity201}
\end{equation}
\begin{equation}
u_{1\alpha}^{(0,0)} = -\frac{k_1(k_2+1)b_{;\alpha}}{(k_1+1)(k_2-2k_1+1)b}.
\label{velocity100}
\end{equation}
The direct check shows that the velocity coefficients (\ref{velocity201}) 
and (\ref{velocity100}) satisfy Eq. (\ref{eq-vel}). 

The next order term of the quasi-isotropic series solution (proportional 
to $t^{1-2\gamma_0}$ in the Einstein equation (\ref{Einstein2}))
immediately gives
\begin{equation}
(1-\gamma_0)(c_{\alpha\ ;\beta}^{\beta}-c_{;\alpha}) = 8\pi G (1+k_2) 
\varepsilon_2^{(0,0)}u_{2\alpha}^{(1,0)}.
\label{eq-vel3}
\end{equation}
Notice that in the absence of degeneracy between exponents of different
terms in the perturbation series, this equation contains the contribution 
from the firstly dominant fluid 2 only.
Thus, using the explicit expression for the coefficients 
$c_{\alpha}^{\beta}$ (see Eq. (\ref{c}) and the Bianchi identity for the 
curvature tensor $\tilde{P}_{\alpha}^{\beta}$
\begin{equation}
\tilde{P}_{\alpha\ ;\beta}^{\beta} = \frac12 \tilde{P}_{;\alpha}~,
\label{BianchiP}
\end{equation}
one has
\begin{equation}
u_{2\alpha}^{(0,0)} = -\frac{27 k_2 (k_2+1)^3 \tilde{P}_{;\alpha}}
{8(3k_2+5)(9k_2+5)}.
\label{velocity210}
\end{equation}

To find next terms of the quasi-isotropic solution for the 
velocities, i.e., corrections proportional to $t^{(3-\gamma_0\alpha_1)}$, 
we shall use the identity (\ref{Bianchi1}) again. This leads to the 
following relations :
\begin{eqnarray}
&&k_2\varepsilon_{2\  ;\alpha}^{(0,2)} = (3-2\gamma_0\alpha_1) 
(\varepsilon_2^{(0,0)}u_{2\alpha}^{(0,2)} 
+\varepsilon_2^{(0,1)}u_{2\alpha}^{(0,1)})
\nonumber \\
&&+\frac12 6\gamma_0(\varepsilon_2^{(0,0)}u_{2\alpha}^{(0,2)} 
+\varepsilon_2^{(0,1)}u_{2\alpha}^{(0,1)})\nonumber \\
&&+\frac12(2-\gamma_0\alpha_1)b\varepsilon_2^{(0,0)}u_{2\alpha}^{(0,1)},
\label{eq-vel4}
\end{eqnarray}
\begin{eqnarray}
&&k_1\varepsilon_{1\  ;\alpha}^{(0,1)} = (3-2\gamma_0\alpha_1) 
(\varepsilon_1^{(0,0)}u_{1\alpha}^{(0,1)} 
+\varepsilon_1^{(0,1)}u_{1\alpha}^{(0,0)})
\nonumber \\
&&+\frac12 6\gamma_0(\varepsilon_1^{(0,0)}u_{1\alpha}^{(0,1)} 
+\varepsilon_1^{(0,1)}u_{1\alpha}^{(0,0)})\nonumber \\
&&+\frac12(2-\gamma_0\alpha_1)b\varepsilon_1^{(0,0)}u_{1\alpha}^{(0,0)}.
\label{eq-vel5}
\end{eqnarray}
 From Eqs. (\ref{eq-vel4}), (\ref{eq-vel5}) one can find the  
formulae for unknown quantities $u_{2\alpha}^{(0,2)}$ and 
$u_{1\alpha}^{(0,1)}$ which read
\begin{eqnarray}
&&u_{2\alpha}^{(0,2)} = \frac{1}{(3k_2-4k_1+1)\varepsilon_2^{(0,0)}}
\times(k_2(k_2+1)\varepsilon_{2\ ;\alpha}^{(0,2)} \nonumber \\
&&-(3k_2-4k_1+1)\varepsilon_2^{(0,1)}u_{2\alpha}^{(0,1)} 
-(k_2-k_1)b\varepsilon_2^{(0,0)}u_{2\alpha}^{(0,1)}),
\label{eq-vel6}
\end{eqnarray}
\begin{eqnarray}
&&u_{1\alpha}^{(0,1)} = \frac{1}{(3k_2-4k_1+1)\varepsilon_1^{(0,0)}}
\times(k_1(k_2+1)\varepsilon_{1\ ;\alpha}^{(0,1)} \nonumber \\
&&-(3k_2-4k_1+1)\varepsilon_1^{(0,1)}u_{1\alpha}^{(0,0)} 
-(k_2-k_1)b\varepsilon_1^{(0,0)}u_{1\alpha}^{(0,0)}).
\label{eq-vel7}
\end{eqnarray}
Substituting the explicit expressions for $\varepsilon_2^{(0,0)},
\varepsilon_2^{(0,1)},\varepsilon_2^{(0,2)},\varepsilon_1^{(0,0)},
\varepsilon_1^{(0,1)},u_{2\alpha}^{(0,1)},u_{1\alpha}^{(0,0)}$
from Eqs. (\ref{energy200}), (\ref{energy201}), (\ref{energy202}), 
(\ref{energy100}), (\ref{energy101}), (\ref{velocity201}), 
(\ref{velocity100}) into Eqs. (\ref{eq-vel6}) and (\ref{eq-vel7}), we get
\begin{eqnarray}
&&u_{2\alpha}^{(0,2)} = \frac{k_2(k_2+1)bb_{;\alpha}}
{4(k_2-2k_1+1)(5k_2-4k_1+1)(3k_2-4k_1+1)}\nonumber \\
&&\times(-9k_2^3 + 18k_1k_2^2 - 14k_1^2k_2 + 4k_1^3 + 3k_2^2\nonumber\\
&&-6k-1k_2 + 2k-1^2 + 5k_2 - 4k-1 + 1),
\label{velocity202}
\end{eqnarray}
\begin{equation}
u_{1\alpha}^{(0,1)} = \frac{k_1^2(k_2+1)(1-k_1)(k_2-2k_1-1)b_{;\alpha}}
{2(k_1+1)(3k_2-4k_1+1)(k_2-2k_1+1)}.
\label{velocity101}
\end{equation}
 
In conclusion, let us write down the expressions for the first terms of 
the quasi-isotropic series solution for the metric, energy densities of 
two fluids and their 3-velocities calculated above:
\begin{eqnarray}
&&\gamma_{\alpha\beta} = a_{\alpha\beta} t^{\frac{4}{3(k_2+1)}} 
+ b a_{\alpha\beta} t^{\frac{2(3k_2-3k_1+2)}{3(k_2+1)}}\nonumber \\
&&-\frac{9(k_2+1)^2}{(3k_2+1)(3k_2+5)}\left(\tilde{P}_{\alpha\beta}
+\frac{(3k_2^2-6k_2-5) \tilde{P}a_{\alpha\beta}}{4(9k_2+5)}\right)
t^2\nonumber \\
&&-\frac{(3k_2^2-6k_1^2+6k_1k_2+2k_2+2k_1+1)b^2a_{\alpha\beta}}
{12(5k_2-4k_1+1)}
t^{\frac{2(6k_2-6k_1+2)}{3(k_2+1)}} + \cdots,
\label{metricf}
\end{eqnarray}
\begin{eqnarray}
&&\varepsilon_2 = \frac{1}{6\pi G(k_2+1)^2t^2} - \frac{b}{12\pi G(k_2+1)} 
t^{-\frac{2(k_1+1)}{k_2+1}}\nonumber \\
&&+ \frac{3(k_2+1)\tilde{P}}{16\pi G(9k_2+5)} 
t^{-\frac{4}{3(k_2+1)}}\nonumber \\
&&+\frac{(3k_2^2-k_1^2-k_1k_1+5k_2-3k_1+1)b^2}
{24\pi G(k_2+1)(5k_2-4k_1+1)}t^{\frac{2(k_2-2k_1-1)}{k_2+1}} + \cdots,
\label{energy2f}
\end{eqnarray}
\begin{eqnarray}
&&\varepsilon_1 = \frac{(3k_2-2k_1+1)b}{12\pi G(k_2+1)^2} 
t^{-\frac{2(k_1+1)}{k_2+1}}\nonumber \\
&&-\frac{(k_1+1) (3k_2-2k_1+1)b^2}{24\pi G(k_2+1)^2} 
t^{\frac{2(k_2-2k_1-1)}{k_2+1}} + \cdots,
\label{energy1f}
\end{eqnarray}
\begin{eqnarray}
&&u_{2\alpha} = -\frac{k_2(k_2+1)b_{;\alpha}}{2(k_2-2k_1+1)}
t^{\frac{3k_2-2k_1+1}{k_2+1}}\nonumber \\
&&- \frac{27k_2(k_2+1)^3\tilde{P}_{;\alpha}}{8(3k_2+5)(9k_2+5)}
t^{\frac{9k_2+5}{3(k_2+1)}}\nonumber \\
&&+\frac{k_2(k_2+1)bb_{;\alpha}}
{4(k_2-2k_1+1)(5k_2-4k_1+1)(3k_2-4k_1+1)}\nonumber \\
&&\times(-9k_2^3 + 18k_1k_2^2 - 14k_1^2k_2 + 4k_1^3 + 3k_2^2\nonumber\\
&&-6k_1k_2 + 2k_1^2 + 5k_2 - 4k_1 + 1)
t^{\frac{5k_2-4k_1+1}{k_2+1}} + \cdots,
\label{velocity2f}
\end{eqnarray}
\begin{eqnarray}
&&u_{1\alpha} = \frac{k_1(k_2+1)b_{;\alpha}}{(k_1+1)(k_2-2k_1+1)b}t 
\nonumber\\
&&+\frac{k_1^2(k_2+1)(1-k_1)(k_2-2k_1-1)b_{;\alpha}}
{2(k_1+1)(3k_2-4k_1+1)(k_2-2k_1+1)}t^{\frac{3k_2-2k_1+1}{k_2+1}} + 
\cdots .
\label{velocity1f}
\end{eqnarray}

Note that the velocity flow is potential. Actually, it can be shown that 
this important property remains in all higher orders of the quasi-isotropic 
solution. In the degenerate case (see Eq. (\ref{case2})) when the relation 
between coefficients $k_1$ and $k_2$ is 
\begin{equation}
k_1 = \frac{3k_2 - 1}{6},
\label{case21}
\end{equation}
the terms with superscripts $(1,0)$ and  $(0,2)$ in the expressions 
(\ref{metricf}), (\ref{energy2f}) and (\ref{velocity2f}), i.e., the terms 
proportional to $\tilde{P}$ and $b^2$ respectively belong to the same 
order in the quasi-isotropic series expansion. Then the coefficients at 
the corresponding powers of time $t$ should be simply added due to 
linearity of all iterative equations used above. 
 
\section{Conclusions and discussion}

The equations (\ref{metricf})-(\ref{velocity1f}) represent the main result
of the present paper -- the first terms in the generalized quasi-isotropic 
series solution for the case of two barotropic fluids. How many arbitrary 
physical (i.e., gauge invariant) functions of spatial coordinates does it 
contain? The answer is four: 3 functions are contained in $a_{\alpha\beta}$ 
(no conditions are imposed on this symmetric tensor, and the remaining
freedom of coordinate transformations at $t=0$ consists of 3 spatial
rotations ) and the fourth function is $b$. The former 3 physical 
functions describe the growing mode of scalar adiabatic perturbations
and the non-decreasing mode of gravitational waves (with 2 polarization
states) in the regime when metric perturbations are not small, while
the latter function describes the growing mode of scalar isocurvature
fluctuations.

It is seen that the first term of the generalized quasi-isotropic 
solution remains factorized with respect to $t$ and spatial coordinates. 
This represents the generalization of the known result, that small 
relative metric perturbations in the synchronous system of reference 
corresponding to the growing adiabatic mode remain constant in the 
long-wave limit even in a multi-fluid case, to the case when metric 
perturbations are not small. For the other, isocurvature mode, metric 
perturbations are zero at $t=0$, but grows with time. However, their back 
reaction appears in higher order terms of the generalized quasi-isotropic 
solution and does not affect the leading term of the adiabatic mode (as 
far as the series expansion has sense at all).

However, as compared to the one-fluid case, the following new feature
arises. In that case, the quasi-isotropic series solution loses sense in 
the course of future expansion when spatial gradients of all quantities 
become comparable to temporal ones. The series for the generalized 
quasi-isotropic solution may diverge for another reason even if spatial
derivatives are still small, namely, when higher-order terms containing
$b$ become of the order of the leading term. This means that the
firstly subdominant fluid becomes dominant and vice versa. Then the
law of expansion should change, too (as occurs, e.g., in the standard
CDM+radiation FRW model after the moment of matter--radiation equality).
This shows that all terms in the generalized quasi-isotropic series 
containing the function $b({\bf r})$ but not its derivatives can be 
summed. We leave this question for future work. 

This project was partially supported by RFBR via grants No 02-02-16817 and 
00-15-96699, by the RAS Research Programme ``Quantum Macrophysics'' 
and by the International Scientific Program ``Relativistic Astrophysics" by the Ministry of Industry, Science and Technology of the Russian Federation.

\end{document}